\documentclass[aps,prl,twocolumn,showpacs,superscriptaddress]{revtex4-2}
\usepackage{amsmath,amssymb}
\usepackage{physics}
\usepackage{hyperref}
\usepackage{graphicx}

\begin{document}

\title{Possible Topological Decoherence Transition in Relativistic Electron Beams\\
Propagating through Coulomb-Disordered Media}
\author{Yury A. Budkov}
\email{ybudkov@hse.ru}
\affiliation{Laboratory of Computational Physics, HSE University, Tallinskaya st. 34, 123458 Moscow, Russia}
\affiliation{Frumkin Institute of Physical Chemistry and Electrochemistry Russian Academy of Sciences, 31-4, Leninsky Prospect, 119071 Moscow, Russia}

\begin{abstract}
We show that the mutual coherence of a relativistic electron beam in a Coulomb-disordered medium is governed by an effective two-dimensional compact phase field with a logarithmic correlation function.  The corresponding Gaussian free-field action exhibits a stiffness inversely proportional to the propagation length.  When the compact nature of the phase is taken into account, the system supports vortex excitations that interact as a two-dimensional Coulomb gas.  Renormalization-group analysis of this gas indicates the existence of a critical sample thickness $L_c$ at which a Berezinskii--Kosterlitz--Thouless (BKT) transition may occur, separating a regime of algebraic decoherence from one where free vortices proliferate and coherence is destroyed exponentially.  The critical thickness is expressed through fundamental microscopic parameters and could be observed in transmission electron microscopy of liquid cells or cryogenic samples.
\end{abstract}

\maketitle

\paragraph{Introduction.}
The recent theory of electron localization and coherence in Coulomb-disordered media~\cite{budkov2026static,budkov2026dynamic,budkov2026coherence} has established a universal relation between the transverse coherence length $\rho_c$ and the single-particle localization length $\ell$,
\begin{equation}
\rho_c \sim \lambda_D \sqrt{\frac{\ell}{L}},
\label{eq:universal}
\end{equation}
where $\lambda_D$ is the Debye screening length and $L$ the sample thickness.  This result links Anderson localization to the intrinsic resolution limit in electron microscopy.  A relativistic extension of the formalism~\cite{budkov2026relativistic} showed that the effective coupling constant saturates at high beam energies, implying that standard TEM voltages already lie near the optimal regime for minimizing Coulomb decoherence.

In the present Letter we go beyond the analysis of the characteristic coherence scale and investigate the asymptotic large-distance behaviour of the mutual coherence function.  We demonstrate that the phase fluctuations of the electron wave front can be mapped onto an effective two-dimensional field theory with compact phase, which supports vortex excitations and may exhibit a BKT-type topological transition as a function of sample thickness.

\begin{figure*}[t]
\centering
\includegraphics[width=\textwidth]{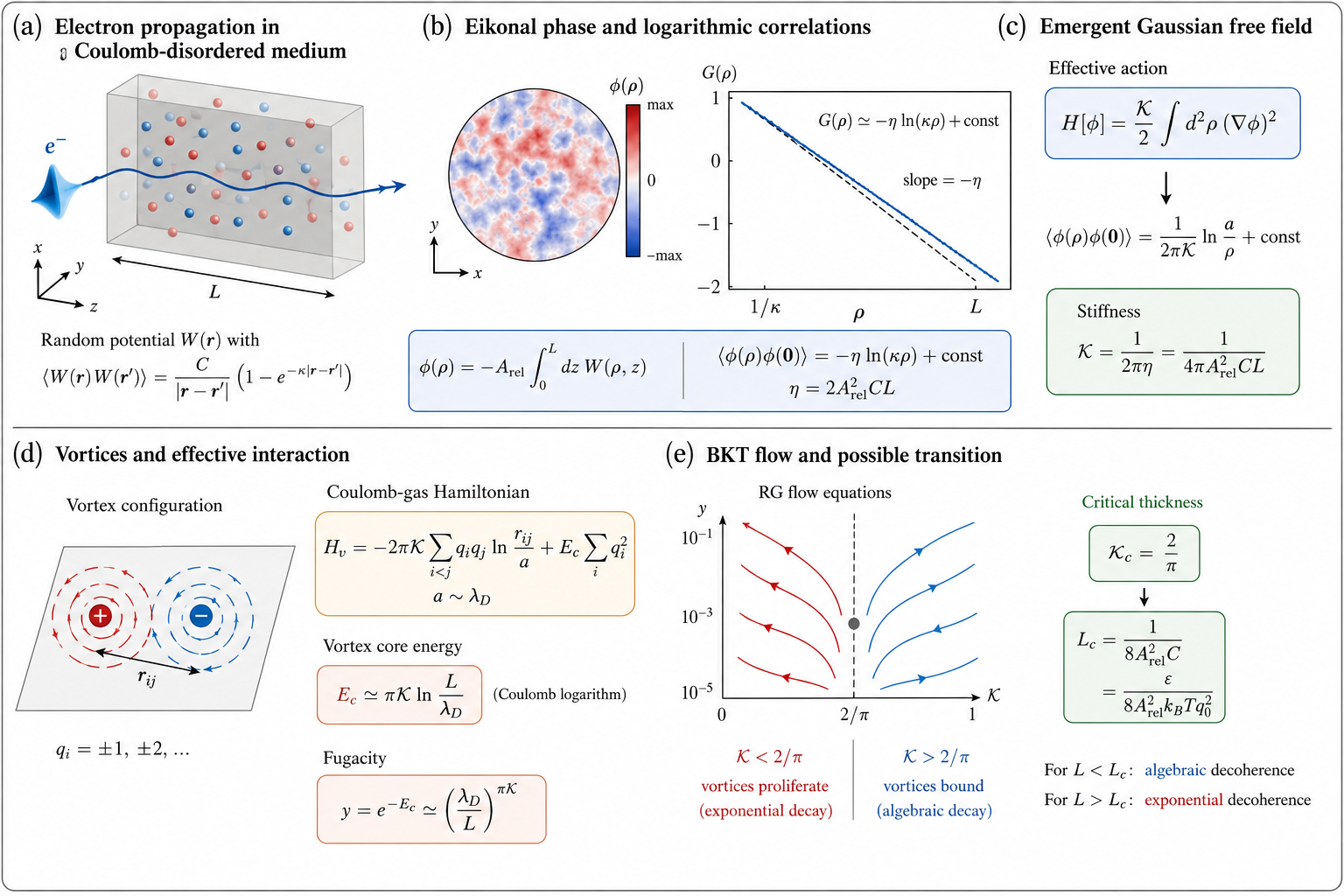}
\caption{
Schematic illustration of the emergence of an effective two-dimensional field theory for electron decoherence in a Coulomb-disordered medium and the possible Berezinskii--Kosterlitz--Thouless (BKT) scenario.
(a) Relativistic electron propagation through a fluctuating Coulomb environment characterized by the screened correlation function
$K(r)=\frac{C}{r}(1-e^{-\kappa r})$.
The accumulated eikonal phase is generated by longitudinal integration of the random potential.
(b) The resulting phase field $\phi(\boldsymbol{\rho})$ exhibits logarithmic correlations,
$\langle\phi(\rho)\phi(0)\rangle \sim -\eta\ln(\kappa\rho)$,
leading to algebraic decay of coherence.
(c) Infrared reduction to a two-dimensional Gaussian free field with effective stiffness
$\mathcal K = (4\pi A_{\rm rel}^2 C L)^{-1}$.
Increasing sample thickness $L$ softens the stiffness and enhances phase fluctuations.
(d) Assuming compactness of the effective phase field, singular vortex configurations become possible.
Their interaction energy maps onto a two-dimensional Coulomb gas with logarithmic interaction potential.
The vortex core energy acquires the form
$E_c \sim \pi \mathcal K \ln(L/\lambda_D)$,
which naturally contains the Coulomb logarithm.
(e) Possible BKT renormalization-group flow.
For $\mathcal K > 2/\pi$, vortices remain bound and coherence decays algebraically,
whereas for $\mathcal K < 2/\pi$ free vortices proliferate, potentially driving exponential decoherence.
The corresponding critical thickness is
$L_c = (8A_{\rm rel}^2 C)^{-1}$.
The figure summarizes the proposed connection between electron decoherence in correlated Coulomb media and emergent two-dimensional topological field theory.
}
\label{fig:BKT}
\end{figure*}

\paragraph{Effective phase dynamics.}
We consider a relativistic electron beam propagating along $z$ through a classical one-component plasma.  After the paraxial reduction of the Dirac equation, the envelope wave function obeys~\cite{budkov2026relativistic}
\begin{equation}
i\partial_z \psi = -\frac{1}{2k}\nabla_\perp^2 \psi + A_{\rm rel} W(\boldsymbol{\rho},z)\psi,
\label{eq:paraxial}
\end{equation}
where $k$ is the longitudinal wave number, $A_{\rm rel}=(\gamma+1)/(2\gamma\hbar v)$ the relativistic coupling constant, and $W$ a static Gaussian random potential with correlator
\begin{equation}
\langle W(\mathbf r) W(0)\rangle = K(r) = \frac{C}{r}\bigl(1-e^{-\kappa r}\bigr).
\label{eq:K}
\end{equation}
Here $C=k_B T q_0^2/\varepsilon$, $\kappa=\lambda_D^{-1}$, and $\varepsilon$ is the dielectric constant.

In the eikonal approximation the phase accumulated over the sample thickness $L$ is $\phi(\boldsymbol{\rho},L) = -A_{\rm rel}\int_0^L dz\,W(\boldsymbol{\rho},z)$.  For $L\gg\lambda_D$ the phase correlator takes the form
\begin{equation}
G(\rho)\equiv\langle\phi(\boldsymbol{\rho})\phi(0)\rangle \simeq A_{\rm rel}^2 L C \bigl[-2\ln(\kappa\rho) + \text{const}\bigr],
\label{eq:G}
\end{equation}
valid for $\kappa\rho\gg1$.  The corresponding phase structure function is $D_\phi(\rho)=2[G(0)-G(\rho)]\simeq 4A_{\rm rel}^2 C L \ln(\kappa\rho)$, leading to an algebraic decay of the mutual coherence,
\begin{equation}
\gamma(\rho) = e^{-D_\phi(\rho)/2} \sim (\kappa\rho)^{-\eta},
\qquad \eta = 2 A_{\rm rel}^2 C L.
\label{eq:gamma}
\end{equation}

\paragraph{Emergent Gaussian free field.}
The logarithmic form of $G(\rho)$ identifies the fluctuating phase as a two-dimensional Gaussian free field with effective Hamiltonian
\begin{equation}
H[\phi] = \frac{\mathcal{K}}{2} \int d^2\rho\, (\nabla\phi)^2,
\label{eq:Hgauss}
\end{equation}
where the stiffness is
\begin{equation}
\mathcal{K} = \frac{1}{2\pi\eta} = \frac{1}{4\pi A_{\rm rel}^2 C L}.
\label{eq:Keff}
\end{equation}
Thus the longitudinal propagation acts as a renormalization parameter: increasing $L$ reduces the transverse phase stiffness.

\paragraph{Compact phase and vortices.}
The physical phase is defined only modulo $2\pi$, i.e.\ $\phi\equiv\phi+2\pi$.  This compactness admits topological excitations--vortices--that can be described by the singular part $\phi_v = \sum_i q_i \arg(\boldsymbol{\rho}-\boldsymbol{\rho}_i)$ with integer charges $q_i$.  Substituting the decomposition $\phi=\phi_s+\phi_v$ into Eq.~\eqref{eq:Hgauss} and integrating out the smooth part yields the vortex Hamiltonian
\begin{equation}
H_v = -2\pi \mathcal{K} \sum_{i<j} q_i q_j \ln\frac{|\boldsymbol{\rho}_i-\boldsymbol{\rho}_j|}{a}
      + E_c \sum_i q_i^2,
\label{eq:Hv}
\end{equation}
where $a\sim\lambda_D$ is the vortex core radius and $E_c\sim 2\pi\mathcal{K}\ln(a/a_0)$ the core energy.  This is the Hamiltonian of a two-dimensional Coulomb gas of charged particles.

\paragraph{Possible BKT transition.}
The grand-partition function of the vortex gas is precisely the starting point of the Kosterlitz--Thouless renormalization-group analysis~\cite{Kosterlitz1973,Berezinskii1972}.  The stiffness $\mathcal{K}$ and the vortex fugacity $y=e^{-E_c}$ are expected to flow according to the standard KT recursion relations.  These equations possess a critical point at $\mathcal{K}_c=2/\pi$, $y=0$, separating a phase of bound vortex pairs (algebraic order) from a phase of free vortices (exponential decay of correlations).  Using the microscopic expression~\eqref{eq:Keff} for the stiffness, this critical condition translates into a critical thickness
\begin{equation}
L_c = \frac{1}{8 A_{\rm rel}^2 C} = \frac{\varepsilon}{8 A_{\rm rel}^2 k_B T q_0^2}.
\label{eq:Lc}
\end{equation}

It must be emphasized that the existence of a BKT transition relies on the dynamical generation of free vortices in the electron problem, which has not been proven microscopically.  Nevertheless, the mathematical structure derived here strongly suggests that such a transition may occur.  For $L<L_c$ the stiffness $\mathcal{K}>\mathcal{K}_c$, vortices are bound, and the algebraic decoherence of Eq.~\eqref{eq:gamma} survives on all scales.  For $L>L_c$, $\mathcal{K}<\mathcal{K}_c$, and free vortices may proliferate, leading to exponential decay of $\gamma(\rho)$ and thus to a qualitatively stronger decoherence.

\paragraph{Experimental implications.}
The predicted critical thickness $L_c$ is expressed in terms of measurable quantities.  For a typical aqueous electrolyte ($\varepsilon=80$, $T=300\,$K, $q_0=e$) and $A_{\rm rel}\approx1/(\hbar v)$ at non-relativistic energies, one estimates $L_c\sim 10^{-5}$--$10^{-4}\,$cm (100--1000~nm).  This scale is well within the range of modern liquid-cell TEM and cryo-EM experiments.  Observing the predicted crossover from algebraic to exponential decoherence as a function of $L$ would provide direct evidence for a BKT-type transition in electron coherence.  Alternatively, varying the ionic strength or the dielectric permittivity of the solvent should shift $L_c$ in a predictable way.

\paragraph{Conclusion.}
We have demonstrated that the Coulomb-disordered electron propagation problem naturally gives rise to an effective two-dimensional compact Gaussian field theory.  The corresponding stiffness is inversely proportional to the sample thickness, and the inclusion of vortex excitations leads to the mathematical structure of a BKT transition.  A critical thickness $L_c$ is identified below which decoherence is algebraic and above which free vortices may drive an exponential loss of coherence.  These findings establish a deep connection between quantum particle localization, wave propagation in random media, and topological phase transitions, and they open a new direction for exploring coherence phenomena in electron microscopy.

\bibliography{relativistic}

\clearpage
\onecolumngrid

\section*{End Matter}

\subsection*{Microscopic model and accumulated eikonal phase}

We consider the relativistic paraxial equation for the slowly varying
electron envelope,
\begin{equation}
i\partial_z \psi(\boldsymbol{\rho},z)
=
-\frac{1}{2k}\nabla_\perp^2\psi
+
A_{\rm rel}W(\boldsymbol{\rho},z)\psi ,
\label{eq:EMparaxial}
\end{equation}
where $k$ is the longitudinal wave number,
\[
A_{\rm rel}
=
\frac{\gamma+1}{2\gamma\hbar v},
\]
and $W(\mathbf r)$ is a static Gaussian random potential satisfying
\begin{equation}
\langle W(\mathbf r)\rangle =0,
\qquad
\langle W(\mathbf r)W(\mathbf r')\rangle
=
K(\mathbf r-\mathbf r').
\end{equation}

Within the random-phase approximation the correlator is taken in the form
\begin{equation}
K(r)
=
\frac{C}{r}\left(1-e^{-\kappa r}\right),
\label{eq:EMK}
\end{equation}
where
\[
C
=
\frac{k_BTq_0^2}{\varepsilon},
\qquad
\kappa=\lambda_D^{-1},
\]
and $\varepsilon$ is the dielectric permittivity.

In the eikonal regime the dominant effect of the random potential is
phase accumulation along the propagation direction. The accumulated
phase at the exit plane $z=L$ is therefore
\begin{equation}
\phi(\boldsymbol{\rho})
=
-A_{\rm rel}\int_0^Ldz\,W(\boldsymbol{\rho},z).
\label{eq:EMphi}
\end{equation}

Since $W$ is Gaussian, the field $\phi(\boldsymbol{\rho})$ is also
Gaussian and is completely characterized by its two-point correlation
function.

\subsection*{Phase correlator and logarithmic structure function}

The phase correlator is
\begin{equation}
G(\rho)
\equiv
\langle
\phi(\boldsymbol{\rho})\phi(\mathbf 0)
\rangle .
\end{equation}

For sufficiently thick samples, $L\gg\lambda_D$, translational invariance
along the propagation direction gives
\begin{equation}
G(\rho)
=
A_{\rm rel}^2L
\int_{-\infty}^{\infty}du\,
K\!\left(\sqrt{\rho^2+u^2}\right).
\label{eq:EMG}
\end{equation}

Substituting Eq.~(\ref{eq:EMK}) yields
\begin{align}
G(\rho)
&=
A_{\rm rel}^2LC
\left[
I_1(\rho)-I_2(\rho)
\right],
\\
I_1(\rho)
&=
\int_{-\infty}^{\infty}
\frac{du}{\sqrt{\rho^2+u^2}},
\\
I_2(\rho)
&=
2K_0(\kappa\rho),
\end{align}
where $K_0$ is the modified Bessel function.

The screened contribution $I_2(\rho)$ decays exponentially for
$\kappa\rho\gg1$, whereas $I_1(\rho)$ possesses the logarithmic
dependence
\begin{equation}
I_1(\rho)
\simeq
-2\ln(\kappa\rho)+\mathrm{const}.
\end{equation}

Consequently,
\begin{equation}
G(\rho)
\simeq
-\eta\ln(\kappa\rho)+\mathrm{const},
\label{eq:EMGlog}
\end{equation}
with
\begin{equation}
\eta
=
2A_{\rm rel}^2CL.
\label{eq:EMeta}
\end{equation}

Although both $G(0)$ and $G(\rho)$ are individually infrared divergent,
the physically observable phase structure function remains finite:
\begin{equation}
D_\phi(\rho)
=
\left\langle
[\phi(\boldsymbol{\rho})-\phi(\mathbf0)]^2
\right\rangle
=
2[G(0)-G(\rho)].
\end{equation}

Using Eq.~(\ref{eq:EMGlog}) one obtains
\begin{equation}
D_\phi(\rho)
\simeq
2\eta\ln(\kappa\rho).
\label{eq:EMDphi}
\end{equation}

The corresponding coherence function therefore decays algebraically:
\begin{equation}
\gamma(\rho)
=
\exp[-D_\phi(\rho)/2]
\sim
(\kappa\rho)^{-\eta}.
\label{eq:EMgamma}
\end{equation}

\subsection*{Emergent infrared Gaussian field theory}

The logarithmic correlator~(\ref{eq:EMGlog}) is characteristic of a
two-dimensional massless Gaussian field. Consider the effective
Euclidean Hamiltonian
\begin{equation}
H[\phi]
=
\frac{\mathcal K}{2}
\int d^2\rho\,
(\nabla\phi)^2 .
\label{eq:EMH}
\end{equation}

Its two-point correlation function is
\begin{equation}
\langle
\phi(\boldsymbol{\rho})\phi(\mathbf0)
\rangle
=
\frac{1}{2\pi\mathcal K}
\ln\frac{a}{\rho}
+\mathrm{const},
\label{eq:EMGff}
\end{equation}
where $a$ is a short-distance cutoff.

Comparing Eqs.~(\ref{eq:EMGlog}) and~(\ref{eq:EMGff}) gives
\begin{equation}
\mathcal K
=
\frac{1}{2\pi\eta}
=
\frac{1}{4\pi A_{\rm rel}^2CL}.
\label{eq:EMKeff}
\end{equation}

Thus the infrared statistics of the accumulated eikonal phase are
effectively described by a two-dimensional Gaussian free field.
Within this representation the propagation length $L$ plays the role
of an inverse effective stiffness: increasing $L$ enhances phase
fluctuations and reduces $\mathcal K$.

An important aspect of the present problem is that this effective
two-dimensional statistical field theory emerges not from equilibrium
thermal fluctuations of an order parameter, but from the accumulated
random eikonal phase generated during propagation through a
three-dimensional Coulomb-disordered medium.

\subsection*{Compact phase and topological configurations}

Observable interference quantities depend on the phase factor
$e^{i\phi}$ rather than on $\phi$ itself. At the level of the
infrared effective theory, the phase field may therefore be regarded
as effectively compact,
\[
\phi\equiv\phi+2\pi .
\]

Such compactification formally admits topological defect
configurations in the effective long-wavelength theory. A vortex
configuration may be represented as
\begin{equation}
\phi_v(\boldsymbol{\rho})
=
\sum_i
q_i
\arg(\boldsymbol{\rho}-\boldsymbol{\rho}_i),
\end{equation}
where $q_i\in\mathbb Z$ are integer winding numbers.

Whether such singular phase textures are dynamically realized in the
underlying microscopic electron problem remains an open question.
Nevertheless, assuming their existence, the resulting effective
description acquires the structure of a two-dimensional Coulomb gas.

\subsection*{Effective vortex interaction}

Using the standard decomposition
\[
\phi=\phi_s+\phi_v,
\]
into smooth and singular contributions and substituting it into
the Gaussian Hamiltonian~(\ref{eq:EMH}), one obtains the effective
interaction between vortex configurations.
For a set of vortices located at positions
$\boldsymbol{\rho}_i$ with integer winding numbers
$q_i\in\mathbb Z$, the singular field is
\[
\phi_v(\boldsymbol{\rho})
=
\sum_i
q_i
\arg(\boldsymbol{\rho}-\boldsymbol{\rho}_i).
\]

Integrating over the smooth component $\phi_s$
gives the standard logarithmic interaction energy
between vortices~\cite{Berezinskii1972,Kosterlitz1973},
\begin{equation}
H_v
=
-2\pi\mathcal K
\sum_{i<j}
q_iq_j
\ln
\frac{
|\boldsymbol{\rho}_i-\boldsymbol{\rho}_j|
}{a}
+
E_c
\sum_i q_i^2 ,
\label{eq:EMHv}
\end{equation}
where $a\sim\lambda_D$ is the short-distance cutoff of the
effective theory.

The energy of an isolated vortex follows from the logarithmic
gradient energy associated with the singular phase texture,
\begin{equation}
E_v
\simeq
\pi\mathcal K
\int_a^R
\frac{dr}{r}
=
\pi\mathcal K
\ln\frac{R}{a},
\label{eq:EMEv}
\end{equation}
where $R$ is the large-distance cutoff.
In the present problem, the natural ultraviolet cutoff is
set by the Debye screening length,
\[
a\sim\lambda_D,
\]
while the infrared cutoff is controlled by the finite propagation
geometry and may be estimated as
\[
R\sim L.
\]

Consequently,
\begin{equation}
E_v
\sim
\pi\mathcal K
\ln\frac{L}{\lambda_D}.
\label{eq:EMEv2}
\end{equation}

This logarithmic factor is formally analogous to the Coulomb
logarithm discussed previously in the context of localization problem~\cite{budkov2026static,budkov2026dynamic},
\[
\ln\Lambda
=
\ln\frac{L}{\lambda_D}.
\]

The corresponding vortex fugacity is therefore
\begin{equation}
y
\sim
\exp\!\left[
-\pi\mathcal K
\ln\frac{L}{\lambda_D}
\right].
\label{eq:EMfugacity}
\end{equation}

Equations~(\ref{eq:EMHv})--(\ref{eq:EMfugacity})
define an effective two-dimensional Coulomb gas
with logarithmic interactions.

\subsection*{Possible BKT scenario}

If vortex configurations are dynamically generated, the effective
theory belongs to the Berezinskii--Kosterlitz--Thouless universality
class~\cite{Berezinskii1972,Kosterlitz1973}.

Within the normalization adopted in Eq.~(\ref{eq:EMH}), the standard
critical stiffness is
\begin{equation}
\mathcal K_c
=
\frac{2}{\pi}.
\end{equation}

Combining this condition with Eq.~(\ref{eq:EMKeff}) yields the
characteristic thickness scale
\begin{equation}
L_c
=
\frac{1}{8A_{\rm rel}^2C}
=
\frac{\varepsilon}
{8A_{\rm rel}^2k_BTq_0^2}.
\label{eq:EMLc}
\end{equation}

For $L\ll L_c$ the algebraic decoherence behavior described by
Eq.~(\ref{eq:EMgamma}) should dominate. If free vortices become
relevant at larger thicknesses, the effective long-distance behavior
may cross over toward exponentially decaying coherence.

\subsection*{Concluding remarks}

The above analysis demonstrates that Coulomb-induced electron
decoherence in the eikonal regime naturally generates an effective
infrared theory with the structure of a two-dimensional Gaussian
field. Under the additional assumption of effective phase
compactification, the theory formally admits vortex-like
configurations and acquires a close correspondence with the
two-dimensional Coulomb gas underlying BKT physics.

At the same time, the existence and microscopic origin of such
topological configurations in the electron problem remain to be
established rigorously. The present construction should therefore
be viewed as an effective-field-theoretical scenario indicating a
possible route toward vortex-mediated decoherence phenomena in
strongly phase-disordered electronic propagation.

\end{document}